\documentstyle[aps,amsfonts,twocolumn,epsfig]{revtex}
\begin{document}
\draft

\title{Multi-Valued Logic Gates for Quantum Computation}
\author{Ashok Muthukrishnan \cite{email} and C. R. Stroud, Jr.}
\address{The Institute of Optics, University of Rochester,
Rochester, New York 14627}
\date{\today}
\maketitle
%
%
\newcommand{\ket}[1]{\mbox{$|#1\rangle$}}
\newcommand{\ketn}[2]{\mbox{$|#1,\ldots,#2\rangle$}}
\newcommand{\bra}[1]{\mbox{$\langle #1|$}}
\newcommand{\Gam}[2]{\mbox{${\bf \Gamma_#1}[Y_#2]$}}
\newcommand{\Gamp}[3]{\mbox{${\bf \Gamma_#1}[P_#2(#3)]$}}
\newcommand{\postscript}[1] {\centerline{\epsfbox{#1}}}
%
%
\begin{abstract}
We develop a multi-valued logic for quantum computing for use in multi-level
quantum systems, and discuss the practical advantages of this approach for
scaling up a quantum computer. Generalizing the methods of binary quantum
logic, we establish that arbitrary unitary operations on any number of
$d$-level systems ($d > 2$) can be decomposed into logic gates that operate on
only two systems at a time. We show that such multi-valued logic gates are
experimentally feasible in the context of the linear ion trap scheme for
quantum computing. By using $d$ levels in each ion in this scheme, we reduce
the number of ions needed for a computation by a factor of $\log_2 d$.
\end{abstract}
\pacs{03.67.Lx, 03.67.-a, 03.65.Bz, 89.80.+h}
%
\section{Introduction}\label{sec-Intro}
Binary logic gates and Boolean algebra play an important role in classical and
quantum theories of computation.  The unit of memory for binary quantum
computation is the \mbox{qu-bit}, a quantum system existing in a linear
superposition of two basis states, labeled \ket{0} and \ket{1}. Any
computation, however large, can be performed using universal logic gates that
operate on a small, fixed number of bits or qubits. In the quantum case, a
unitary transformation of any number of entangled qubits can be constructed
from logic gates that operate on only two qubits at a time
\cite{DiVincenzo95,Sleator95,Barenco95a,Barenco95b}, a result that has no
analog in classical reversible logic where three-bit gates are needed to
simulate all reversible Boolean functions \cite{Toffoli80}. We consider the
extension of universal quantum logic to the multi-valued domain, where the unit
of memory is the \mbox{qu-dit} \cite{Gottesman99}, a $d\/$-dimensional quantum
system with the basis states, \ket{0}, \ket{1}, \ldots, \ket{d-1}. This offers
greater flexibility in the storage and processing of quantum information, and
more importantly, provides an alternate route to the scaling up of quantum
computation.

As in the binary case, a tensor product of many such qudits is essential for
the efficient storage of information, since the number of dimensions in the
Hilbert space scales exponentially with the number of qudits in the system.
Allowing $d$ to be arbitrary enables a trade-off between the number of qudits
making up the quantum computer and the number of levels in each qudit. For
example, the linear ion trap quantum computer \cite{Cirac95} uses only two
levels in each ion for computing, although additional levels can be accessed,
and are typically needed, for processing and reading out the state of the ion
\cite{Monroe95}. By using $d$ computational levels in each ion, we reduce the
number of ions needed for a computation in this scheme by a factor of
$\log_{2}d$, since the Hilbert space of $n$ qudits has the same dimensionality
as $n(\log_{2}d)$ qubits, namely $d^n = 2^{n \log_{2}d}$. Given the difficulty
of trapping and coherently manipulating a large number of ions in their
vibrational ground state in this scheme, a reduction in the number of ions
offered by a multi-valued memory is an advantage.

A tensor product of qudits is also essential for the efficient processing of
quantum information. As in the binary case, we build unitary transforms on the
whole system from logic gates that operate within and between qudits, creating
entangled superpositions, rather than by transforming subsets of a
non-entangled, unary Hilbert space. These elementary multi-valued gates are
necessarily more complex than their binary counterparts, involving a controlled
transformation of all the levels of each qudit. However, a logical network of
these gates becomes simpler at this expense, invoking a trade-off between the
complexity of each gate and the number of gates needed for a computation
\cite{Rine}. We implement a multi-level gate in the linear ion trap scheme by
using multiple lasers to address the different transitions in each ion
simultaneously. In this approach, each multi-level gate takes less time to
implement than the equivalent binary gate sequence on two-level systems,
enabling larger computations within the decoherence time.

Quantum computing in multi-level systems is ideally described using a
multi-valued basis for logic. The information stored in a $d\/$-level quantum
system is fundamentally non-binary in character, since a measurement collapses
the system to one of these $d$ levels, specifying a single value for the qudit,
rather than the $\log_{2}d$ values characteristic of a binary representation of
the same Hilbert space. Moreover, as the $d$ levels in a single qudit need not
contain any entanglement, two-bit conditional logic among these levels is not
well-defined, and cannot simulate multi-level unitary transforms in practice.
By contrast, the entanglement between two different $d\/$-level systems enables
conditional two-qudit logic gates, which we show to be the elementary
operations of multi-valued quantum computing.

Quantum error-correction codes have recently been extended to the multi-valued
domain, for correcting errors in a single qudit \cite{Chau97}, and multiple
qudits \cite{Rains99,Ashikhmin00}. Fault-tolerant procedures for implementing
two-qudit and three-qudit analogs of universal binary gates have also been
developed \cite{Gottesman99,Aharonov99}. A proposal for using a correlated
photon pair to represent the ternary analog of a qubit has been investigated
\cite{Burlakov99}, but no general scheme for implementing multi-valued quantum
logic has been proposed. In the following two sections, we derive a set of one-
and two-qudit gates that are sufficient for universal multi-valued computing,
and show that these can be implemented using multi-level ions in the linear ion
trap model.

\section{Multi-valued Logic Gates}\label{sec-Gates}

We review the gates that are universal for binary quantum logic in a way that
facilitates their multi-valued generalization. The universal binary gates
belong to a family of unitary transforms described by three parameters
\cite{Barenco95a}. This derives from the fact that up to an overall phase
factor, any two-dimensional unitary matrix can be written as \cite{Barenco95b}
\begin{equation}
\label{Y2}
    Y_2(\lambda, \nu, \phi) \; = \;
\left[
    \begin{array}{c@{\hspace{1.3ex}}c}
        \cos \lambda  &  -\,e^{i\nu} \sin \lambda \\*[0.7ex]
        e^{i(\phi-\nu)} \sin \lambda  &  \ \ e^{i\phi} \cos \lambda
    \end{array}
\right],
\end{equation}
expressed in the basis states of a qubit, \ket{0} and \ket{1}. The parameters
$\lambda$, $\nu$, and $\phi$ are usually taken to be irrational multiples of
$\pi$ and of each other \cite{Barenco95a}, since this allows even a single gate
in Eq.~(\ref{Y2}) to generate all single-qubit transforms asymptotically by
repeated application. However, we find it more useful to consider these three
parameters as arbitrary variables in a simulation, with $Y_2$ representing a
family of gates that can be implemented by appropriate choice of three physical
controls. One of the properties of $Y_2$ is that it can transform any known
state of a qubit to \ket{1}. Such a transformation, labeled $Z_2$, depends on
the coefficients of the state being transformed,
\begin{equation}
\label{Z2}
    \begin{array}{c} \hspace{1ex}
    Z_2(c_0,c_1) = Y_2(\cos^{-1}\! |c_1|, \, \arg[c_0 c_1^*], \, \arg[c_1^*]):
    \\*[1.5ex] \hspace{-1ex}
    c_0\ket{0} + c_1\ket{1} \; \mapsto \; \ket{1},
    \end{array}
\end{equation}
where $|c_0|^2 + |c_1|^2 = 1$. $Y_2$ also contains the phase gate, $X_2$, that
advances the phase of \ket{1} without affecting \ket{0},
\begin{equation}
\label{X2}
    X_2(\phi) = Y_2(0,0,\phi): \;
    \left\{
    \begin{array}{l}
    \ket{1} \,\mapsto\, e^{i\phi} \ket{1};
    \\*[0.4ex]
    \ket{0} \,\mapsto\, \ket{0}.
    \end{array}
    \right.
\end{equation}
Using these two transformation properties of $Y_2$, we can show that the
two-qubit gates that are universal for quantum logic take the form,
\begin{equation}
\label{Gamma22}
    \Gam{2}{2} \; = \;
\left[
    \begin{array}
    {@{\hspace{+2.4ex}}l@{\hspace{+2.4ex}}|@{\hspace{+2.6ex}}c@{\hspace{+1.5ex}}}
        \ & \                   \\*[-0.5ex]
        \hat{1}_2 & \hat{0}       \\*[-0.7ex]
        \ & \                   \\*[-0.5ex]
        \hline
        \ & \                   \\*[-0.6ex]
        \hat{0} & \,Y_2         \\*[-1.5ex]
        \ & \
    \end{array}
\right],
\end{equation}
acting in the four-dimensional basis of the two qubits. The two-dimensional
identity $\hat{1}_2$ acts in the basis of \ket{0,0} and \ket{0,1}, and $Y_2$
acts in the basis of \ket{1,0} and \ket{1,1}. Taken together, this transforms
the second qubit by $Y_2$ conditional on the first qubit being in \ket{1}. The
family of gates, \Gam{2}{2}, is universal for binary quantum logic in the sense
that a unitary transform on any number of qubits can be simulated by repeated
application of these gates on no more than two qubits at a time.

We generalize Eqs.~(\ref{Z2}-\ref{Gamma22}) to the multi-valued case. We define
$Z_d$ as a family of $d$-dimensional transforms that maps a known single-qudit
state to \ket{d-1},
\begin{equation}
\label{Zd}
    \begin{array}{c}
    Z_d(c_0,c_1, \ldots ,c_{d-1}): \\*[+1.2ex]
    c_0\ket{0} + c_1\ket{1} + \cdots + c_{d-1}\ket{d-1} \;
    \mapsto \; \ket{d-1},
    \end{array}
\end{equation}
where the $d$ complex coefficients, $c_0, \ldots, c_{d-1}$, are normalized to
unity, yielding $2d-1$ real quantities that parametrize $Z_d$. As in the binary
case, Eq.~(\ref{Zd}) does not determine $Z_d$ uniquely, since it gives the
transformation of only one of the $d$ states in the basis, namely the
superposition state with the coefficients, $c_0, \ldots, c_{d-1}$. Since it
reduces this superposition to a single specified state, \ket{d-1}, we may
regard $Z_d$ as an instance of the quantum search algorithm \cite{Grover97},
and relate it asymptotically to the Walsh-Hadamard and phase transforms used in
this algorithm. Generalizing Eq.~(\ref{X2}), we define the $d$-dimensional
phase gate $X_d$ as a function of a single parameter,
\begin{equation}
\label{Xd}
    X_d(\phi): \;
    \left\{
    \begin{array}{l}
    \ket{d-1} \,\mapsto\, e^{i\phi}\/\ket{d-1};
    \\*[0.5ex]
    \ket{p} \mapsto \ket{p}
        \mbox{\ for $p \neq d-1$},
    \end{array}
    \right.
\end{equation}
which advances the phase of \ket{d-1} by $\phi$ without affecting any other
state in the qudit. It turns out that the gates, $Z_d$ and $X_d$, are
sufficient to simulate all single-qudit unitary transforms. We can implement
these gates by controlling only $2d$ real parameters, rather than the $d^2-1$
that correspond to generalizing Eq.~(\ref{Y2}), thus greatly simplifying the
physical realization of single-qudit gates. If $Y_d$ represents either $Z_d$ or
$X_d$, then the multi-valued analog of ${\bf \Gamma_2}$ becomes
\begin{equation}
\label{Gamma2d}
    \Gam{2}{d} =
\left[
    \begin{array}
    {@{\hspace{+6ex}}l@{\hspace{+5ex}}|@{\hspace{+2.7ex}}c@{\hspace{+2.1ex}}}
        \ & \       \\*[+4ex]
        \hat{1}_{d^2-d} & \hat{0}   \\*
        \ & \       \\*[+3.8ex]
        \hline
        \ & \       \\*[+0.2ex]
        \hat{0} & \;Y_d   \\*[-0.8ex]
        \ & \
    \end{array}
\right],
\end{equation}
acting in the $d^2$-dimensional basis of two qudits. The identity
$\hat{1}_{d^2-d}$ acts on the states, \ket{0,0}, \ldots, \ket{d-2,d-1}, and
$Y_d$ acts on the remaining $d$ states, \ket{d-1,0}, \ldots, \ket{d-1,d-1}.
This transforms the second qudit by $Y_d$ conditional on the first qudit being
in \ket{d-1}. We now show that such gates are sufficient for constructing
arbitrary unitary transforms on any number of qudits.

Consider an $N$-dimensional unitary transform ${\bf U}$ acting on $n = \log_d
N$ qudits. Each state in the computational Hilbert space can be written as a
tensor product of these $n$ qudits,
\begin{eqnarray}
\label{compstate}
    & \ \ \ket{k} = \ket{k_1}\ket{k_2}\ldots\ket{k_n}, &
\\*[0.5ex]
    \ \ \ & k = 0, 1, \ldots, N-1; \ \
        k_i = 0, 1, \ldots, d-1 \mbox{ for all $i$}, &
\nonumber
\end{eqnarray}
where $k_1\/k_2\/\ldots\/k_n$ is the base-$d$ representation of $k$, with
\ket{k_i} denoting the state of the $i^{\rm{th}}$ qudit.  We will use the
abbreviation, \ketn{k_1,k_2}{k_n}, for \ket{k_1}\ket{k_2}\ldots\ket{k_n}. Let
the eigenstates of ${\bf U}$ be \ket{\Psi_m}, for $m = 1,2, \ldots, N$, with
corresponding eigenvalues $e^{i\Psi_m}$. Each such eigenstate can be expanded
in the computational basis,
\begin{eqnarray}
\label{eigenstate}
    \ket{\Psi_m} & = & c_0\ket{0} + \cdots + c_{N-1}\ket{N-1} \\*
    \ & = &
    c_0\ketn{0}{0} + \cdots + \/c_{N-1}\ketn{d-1}{d-1},
        \nonumber
\end{eqnarray}
where the coefficients are determined by ${\bf U}$. Following an argument given
by Deutsch \cite{Deutsch89}, we write ${\bf U}$ as a product of $N$ unitary
transforms, each $N$-dimensional, that has the same eigenstates and eigenvalues
as that of ${\bf U}$,
\begin{equation}
\label{Usimulate}
    {\bf U} \;=\;
    \sum_{m=1}^{N} e^{i\Psi_m}\ket{\Psi_m}\langle\Psi_m|
    \;=\; {\bf W_1 W_2} \ldots {\bf W_N},
\end{equation}
\vspace{-2ex}
\begin{equation}
\label{Wdef}
    {\bf W_m}: \;
    \left\{
    \begin{array}{lll}
    \ket{\Psi_m} & \!\mapsto\! & e^{i\Psi_m}\ket{\Psi_m}; \\*[0.4ex]
    \ket{\Psi_{m'}} & \!\mapsto\! & \ket{\Psi_{m'}}
        \mbox{\ for $m'\neq m$}.
    \end{array}
    \right.
\end{equation}
The problem then reduces to simulating ${\bf W_m}$ for an arbitrary $m$. We
decompose ${\bf W_m}$ into two transforms that are easier to simulate using
elementary gates,
\begin{equation}
\label{Wsimulate}
    {\bf W_m} = {\bf Z_{m}^{\dagger}\,X_m \,Z_m} = {\bf Z_{m}^{-1}\,X_m \,Z_m},
\end{equation}
where ${\bf Z_m}$ and ${\bf X_m}$ are the $N$-dimensional analogs of $Z_d$ and
$X_d$. We require only that ${\bf Z_m}$ transform the $m^{\rm{th}}$ eigenstate
to \ket{N-1},
\begin{equation}
\label{ZN}
    {\bf Z_m}(c_0, c_1, \ldots, c_{N-1}): \; \ket{\Psi_m} \mapsto \ket{N-1},
\end{equation}
which does not determine the transform uniquely, as in the case of $Z_d$. We
define ${\bf X_m}$ as the transform that advances the phase of \ket{N-1} by the
$m^{\rm{th}}$ eigenphase, leaving all other computational states unchanged,
\begin{equation}
\label{XN}
    {\bf X_m}(\Psi_m): \;
    \left\{
    \begin{array}{l}
    \ket{N-1} \,\mapsto\, e^{i\Psi_m}\/\ket{N-1}, \\*[0.5ex]
    \ket{m'} \,\mapsto\, \ket{m'}
        \mbox{\ for $m' \neq N-1$}.
    \end{array}
    \right.
\end{equation}
We need to show that ${\bf W_m = Z_{m}^{-1}\,X_m\,Z_m}$ satisfies
Eq.~(\ref{Wdef}). First note that
\begin{equation}
\label{Wsim1}
    {\bf Z_{m}^{-1}\,X_m \,Z_m} \ket{\Psi_m} =
    {\bf Z_{m}^{-1}} e^{i\Psi_m} \ket{N-1} =
    e^{i\Psi_m} \ket{\Psi_m}.
\end{equation}
For $m'\neq m$, the state ${\bf Z_m}\ket{\Psi_{m'}}$ has no projection along
\ket{N-1},
\begin{displaymath}
    \langle N-1|{\bf Z_m}|\Psi_{m'}\rangle =
    \langle \Psi_m |{\bf Z_{m}^{\dagger}Z_m}|\Psi_{m'}\rangle =
    \langle \Psi_m | \Psi_{m'} \rangle = 0,
\end{displaymath}
which implies that ${\bf X_m}$ has no effect on ${\bf Z_m}\ket{\Psi_{m'}}$.
Hence,
\begin{equation}
\label{Wsim2}
    {\bf Z_{m}^{\dagger}\,X_m \,Z_m} \ket{\Psi_{m'}} =
    {\bf Z_{m}^{\dagger}\,Z_m} \ket{\Psi_{m'}} =
    \ket{\Psi_{m'}}.
\end{equation}
Combining Eqs.~(\ref{Usimulate}-\ref{Wsimulate}), we see that ${\bf Z_m}$ and
${\bf X_m}$ are sufficient to simulate ${\bf U}$. For $N = d$, this implies
that $Z_d$ and $X_d$ contain all single-qudit unitary transforms. In the
multi-qudit case, we show that ${\bf Z_m}$ and ${\bf X_m}$ can be built from
the elementary two-qudit gates, ${\bf \Gamma_2}[Z_d]$ and ${\bf
\Gamma_2}[X_d]$, and their one-qudit counterparts, $Z_d$ and $X_d$. We first
show this for the $n$-qudit analog of ${\bf \Gamma_2}$,
\begin{equation}\label{Gammand}
    \Gam{n}{d} \,\equiv\,
\begin{array}[t]{l}
        \mbox{Apply $Y_d$ to the $n^{\rm{th}}$ qudit if and only}
        \\*[0.1ex]
        \mbox{if the first $n-1$ qudits are in \ket{d-1},}
\end{array}
\end{equation}
where $Y_d = Z_d$ or $X_d$. Eq.~(\ref{Gammand}) has a matrix representation
analogous to Eq.~(\ref{Gamma2d}), with $\hat{1}_{d^2-d}$ replaced by
$\hat{1}_{d^n-d}$. It is easy to see that ${\bf X_m} = {\bf
\Gamma_n}[X_d(\Psi_m)]$, since ${\bf X_m}$ affects only the last computational
state, $\ket{N-1} = \ketn{d-1}{d-1}$. It is less apparent that ${\bf Z_m}$ is
also contained in Eq.~(\ref{Gammand}). ${\bf Z_m}$ and $Z_d$ are similar in
their transformation properties in that both take a superposition to a single
state. However, $Z_d$ acts within the state space of a single qudit, while
${\bf Z_m}$ transforms the Hilbert space of all $n$ qudits. This suggests that
${\bf Z_m}$ can be achieved by using ${\bf \Gamma_n}[Z_d]$ to target the last
qudit repeatedly, while successively permuting these states with the rest of
the states in the computational basis. First, applying ${\bf
\Gamma_n}[Z_d(c_{N-d},\ldots,c_{N-1})]$ to \ket{\Psi_m} reduces the
superposition of the last $d$ states in Eq.~(\ref{eigenstate}) to \ket{N-1},
shown symbolically as
\begin{eqnarray}
        & \left\{\hspace{0.3ex}
        \ketn{d-1}{d-1,0},\ldots,\ketn{d-1}{d-1,d-1}
        \hspace{-0.3ex}\right\} &
\nonumber \\* \label{Zsim1}
    & =\, \left\{\hspace{0.1ex}\ket{N-d},\ldots,\ket{N-1}\right\}\
        \Rightarrow\ \ket{N-1}. &
\end{eqnarray}
The superposition of the next $d-1$ states in \ket{\Psi_m} is reduced to
\ket{N-1} by permuting these with the last $d$ states but \ket{N-1},
\begin{equation}\label{Zsim2}
\begin{array}{c}
     \left\{\hspace{0.1ex}\ket{N-2d+1},\ldots,\ket{N-d-1}\right\} \\*[0.7ex]
     \Leftrightarrow\
     \left\{\,\ket{N-d},\ldots,\ket{N-2}\hspace{-0.3ex}\right\},
\end{array}
\end{equation}
and using Eq.~(\ref{Zsim1}) again. Continuing in this manner, successive blocks
of $d-1$ states in \ket{\Psi_m} are permuted with the last $d$ states but
\ket{N-1}, and reduced to \ket{N-1} using ${\bf \Gamma_n}[Z_d]$, until the
entire $N$-dimensional state, \ket{\Psi_m}, has been so reduced, completing the
simulation of ${\bf Z_m}$. The permutation of states in Eq.~(\ref{Zsim2}) can
also be done using ${\bf \Gamma_n}[Z_d]$ and ${\bf \Gamma_n}[X_d]$. To see
this, note that a single-qudit permutation is already contained in $Z_d$ and
$X_d$. In particular, if $P_d(p,q)$ denotes the permutation of \ket{p} and
\ket{q} for $p,q = 0, 1, \ldots, d-1$, then
\begin{eqnarray}
\label{permutation1}
    & P_d(p,q) = Z_d^{\dagger}(c_0, \ldots, c_{d-1}) \ X_d(\pi)\
                  Z_d(c_0, \ldots, c_{d-1}), &
\\* \nonumber
    & c_p = - c_q = \frac{1}{\sqrt{2}}\,; \ c_{r \neq p,q} = 0. &
\end{eqnarray}
Permuting two states in the $n$-qudit computational basis,
\begin{eqnarray}
\label{permutation}
    & \ \ \ \ \ketn{j_1,j_2}{j_n}\ \Leftrightarrow\ \ketn{k_1,k_2}{k_n}; &
    \\*[0.2ex] \nonumber
    & \ \ \ \ j_i, k_i = 0, 1, \ldots, d-1 \mbox{ for all $i$}, &
\end{eqnarray}
can be done one qudit at a time, starting with the first qudit.  The last $n-1$
qudits in \ketn{j_1,j_2}{j_n} are first converted to \ket{d-1} by applying a
single-qudit permutation, $P_d(j_i,d-1)$, to each qudit $i$. Then, conditional
\linebreak

\begin{figure}[ht]
\postscript{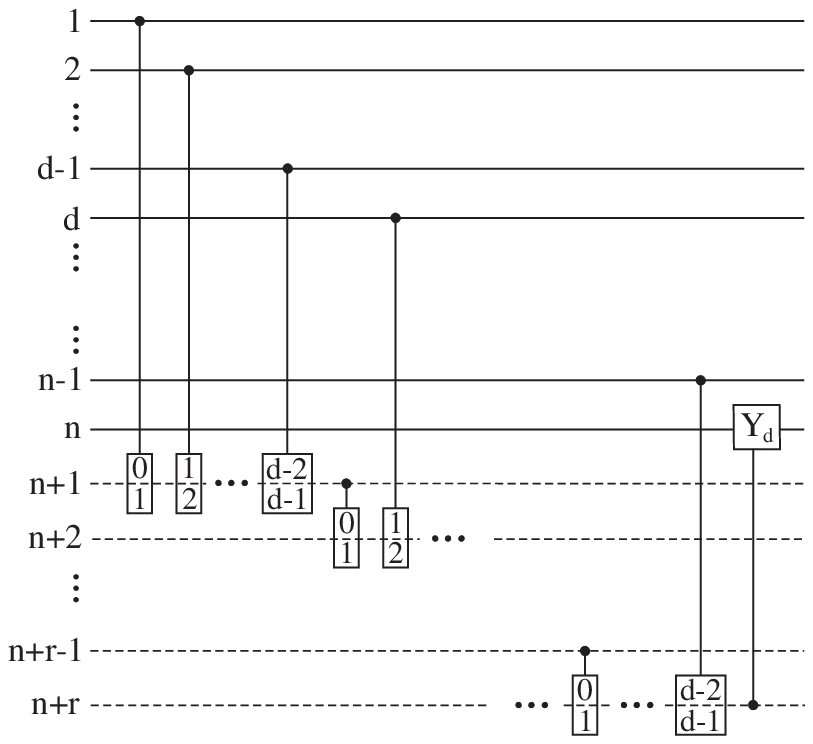}
\vspace{2ex}
\caption{Construction of \Gam{n}{d} from \Gam{2}{d}, for $d>2$.}
\label{gates}
\end{figure}
\vspace{1.8ex}

\noindent on all but the first qudit being in \ket{d-1}, an analog of
\Gamp{n}{d}{j_1,k_1} is applied to permute the first qudit from \ket{j_1} to
\ket{k_1}. The remaining qudits are then restored to their original states by
the same single-qudit permutations, $P_d(j_i,d-1)$, used earlier. This
procedure,
\begin{displaymath}
    \begin{array}{l}
    \prod_{i=2}^n P_d(j_i,d-1):
    \\*[0.8ex]
        \ \ \ \ \ketn{j_1,j_2}{j_n}
        \,\mapsto\,
        \ketn{j_1,d-1}{d-1};
    \\*[1.8ex]
    \Gamp{n}{d}{j_1,k_1}:
    \\*[0.8ex]
        \ \ \ \ \ketn{j_1,d-1}{d-1}
        \,\mapsto\,
        \ketn{k_1,d-1}{d-1};
    \\*[1.8ex]
    \prod_{i=2}^n P_d(j_i,d-1):
    \\*[0.8ex]
        \ \ \ \ \ketn{k_1,d-1}{d-1}
        \,\mapsto\,
        \ketn{k_1,j_2}{j_n},
    \end{array}
\end{displaymath}
is repeated for each of the $n$ qudits, permuting \ketn{j_1,j_2}{j_n} and
\ketn{k_1,k_2}{k_n} without affecting any other computational state.

Thus, any $n\/$-qudit unitary operator ${\bf U}$ can be written in terms of the
logic gates, \Gam{n}{d}, for $Y_d = Z_d$ or $X_d$. We now show that \Gam{n}{d}
can be built from the two-qudit gates, \Gam{2}{d}, of Eq.~(\ref{Gamma2d}). One
way of doing this is illustrated in Fig.~\ref{gates} for $d>2$. The horizontal
lines denote the qudits, with solid lines denoting the $n$ computational qudits
and dashed lines denoting additional auxiliary qudits that have been
initialized to \ket{0}. This simulation uses $r = \lceil(n-2)/(d-2)\rceil$
auxiliary qudits ($\lceil x\rceil$ means the smallest integer greater than
$x$), where $(n-2)/(d-2)$ has been assumed for simplicity to be an integer in
the figure. The vertical lines represent the two-qudit conditional gates,
originating from the control qudit (which is required to be in \ket{d-1} for
the gate to apply) and terminating in a box on the target qudit. The boxes with
two rows p and q represent \Gamp{2}{d}{p,q}, the conditional permutation of
\ket{p} and \ket{q}. The box containing ${\rm{Y}}_{\rm{d}}$ represents
\Gam{2}{d}, for $Y_d = Z_d$ or $X_d$. We want the combination of all these
gates to implement \Gam{n}{d}, applying $Y_d$ to qudit n if and only if the
first $n-1$ qudits are in \ket{d-1}.

Reading from left to right in the figure, the first permutation,
\Gamp{2}{d}{0,1}, increments auxiliary qudit~n+1 from \ket{0} to \ket{1} if and
only if qudit~1 is in \ket{d-1}. The second permutation, \Gamp{2}{d}{1,2},
increments qudit~n+1 from \ket{1} to \ket{2} if and only if qudit~2 is in
\ket{d-1}, and so on. Continuing this way, we see that qudit~n+1 reaches
\ket{d-1} if and only if all of the first $d-1$ computational qudits are in
\ket{d-1}. This information is then transferred to the second auxiliary qudit,
n+2, by the gate \Gamp{2}{d}{0,1}, which increments qudit~n+2 from \ket{0} to
\ket{1} provided qudit~n+1 is in \ket{d-1}. This procedure is carried out
sequentially through all of the computational states, until finally we have the
auxiliary qudit~n+r reaching the state \ket{d-1} (in the case where
$(n-2)/(d-2)$ is an integer) if and only if all of the first $n-1$
computational qudits are in \ket{d-1}. Controlled by this last qudit,
\Gam{2}{d} then acts on qudit~n, completing the simulation of \Gam{n}{d}.
Although not shown in the figure, the two-qudit permutation gates
\Gamp{2}{d}{p,q} are re-applied to the auxiliary qudits at the end to
disentangle them from the computational basis and restore them to \ket{0} for
re-use.

This completes the proof that two-qudit gates of the form, ${\bf
\Gamma_2}[Z_d]$ and ${\bf \Gamma_2}[X_d]$, together with the one-qudit gates,
$Z_d$ and $X_d$, are universal for quantum computing.

\vspace{0.3ex}
%
\section{Ion Trap Implementation}\label{sec-Iontrap}

In this section, we discuss one method of implementing the gates, ${\bf
\Gamma_2}[Z_d]$ and ${\bf \Gamma_2}[X_d]$, in which each qudit is represented
by a $d$-level atom. We use the linear ion trap scheme for quantum computing,
proposed by Cirac and Zoller \cite{Cirac95}, to model the two-qudit
interaction.

The transform $X_d$ does not affect the populations of the $d$ states in the
qudit, but only changes the phase of \ket{d-1}, relative to the other states.
Since only one state is affected in the process, $X_d$ is effectively the same
as its binary counterpart, $X_2$, from a physical standpoint. We can implement
this transform in the atom by coupling \ket{d-1} to an auxiliary state in the
atom using a $2\pi$-pulse, which does not leave any population in this state at
the end of the pulse. In the interaction picture, the phase of \ket{d-1} after
the pulse will be different if the detuning is made time-dependent. One way to
realize a time-dependent detuning is by using a Stark field to shift the
energies of the two levels over time. The other computational states in the
atom are not affected in the process if they are far off-resonance and the
fields are sufficiently weak.

Unlike $X_d$, the transform $Z_d$ involves all of the states in the qudit,
acting on a $d$-state superposition with the coefficients, $c_0, c_1, \ldots,
c_{d-1}$, as shown in Eq.~(\ref{Zd}). The implementation of such a transform
can be posed as a problem in quantum optimal control \cite{Shi90}. If there are
\linebreak

%
\begin{figure}[t]
\postscript{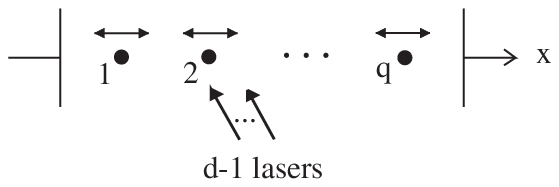}
\caption{Linear ion trap, with $q$ ions. Trap axis is along $x$.}
\label{iontrap}
\end{figure}
\vspace{-3.5ex}
%
%
\begin{figure}[b]
\postscript{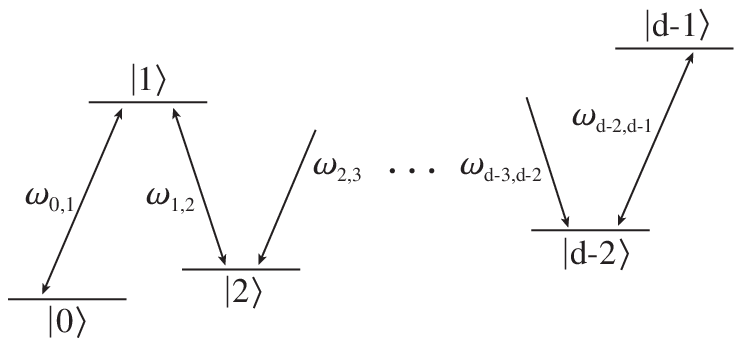}
\vspace{3ex}
\caption{Level scheme for a $d$-level ion, with $d-1$ neighboring transitions.}
\label{levels}
\end{figure}
%
%
\vspace{1.8ex}

\noindent $2d-1$ physical controls
available for manipulating the qudit, an optimization on these controls can be
done with the fidelity governed by Eq.~(\ref{Zd}).

A time-domain approach to this problem in a multi-level atom was studied by
Noel and Stroud \cite{Noel97}, where the control parameters were the amplitudes
and time delays of a sequence of $d$ laser pulses, and the goal was to excite a
$d$-state wave packet in the atom, starting from the ground state. However,
when the ground state population becomes significantly depleted, non-iterative
methods may not be sufficient for creating arbitrary wave packets in the atom
\cite{Araujo98}. We consider an alternate frequency-domain approach to
implementing $Z_d$, where the control parameters are the amplitudes and phases
of different laser fields that are tuned near resonance to $d-1$ atomic
transitions, and that are adiabatically turned on and off over a controlled
time period. To realize the two-qudit gates ${\bf \Gamma_2}$, we consider a
multi-valued extension of the linear ion trap scheme.

Consider $q$ identical $d\/$-level ions confined in a linear harmonic trap with
frequency $\nu_x$, each of which can interact with $d-1$ lasers at a given time
(see Fig.~\ref{iontrap}). If each laser is detuned from the associated atomic
resonance by $\nu_x$, only the center-of-mass (CM) normal mode of the trap is
excited in the absence of power broadening. By laser cooling, the ions are
initially assumed to be in the vibrational ground state of this mode, where
each ion vibrates about its equilibrium position with an amplitude that is
small compared to an optical wavelength. The trap is then characterized by a
Lamb-Dicke parameter, $\eta = k_{x} (\hbar/2m\nu_x)^{\small{1/2}}$, that is
small compared to unity, where $m$ is the mass of each ion and $k_x$ is the
laser wave vector along the trap axis. If $\hat{a}^{\dagger}$ and $\hat{a}$ are
the creation and annihilation operators for the CM mode, and $\hat{\sigma}_{jj}
= \ket{j}\langle j|$ are the internal projection operators for a given
$d\/$-level ion in the trap, the Hamiltonian for this ion in the absence of
interaction fields is \vspace{-0.3ex}
\begin{equation}
\label{hamiltonianbare}
    \hat{H}_0 =\,
    \hbar \nu_x (\hat{a}^{\dagger}\hat{a} + \mbox{$\frac{1}{2}$}) \,+
    \sum_{j=0}^{d-1} \hbar\/ \omega_j\, \hat{\sigma}_{jj}. \vspace{-1ex}
\end{equation}
The computational level scheme considered is shown in Fig.~\ref{levels}, where
the transition frequencies, $\omega_{j,j+1} =$
\mbox{$|\omega_{j+1}-\omega_{j}|$}, are distinct compared to the linewidths of
the levels. For the purpose of implementing $Z_d$, it is sufficient to have
only the neighboring levels coupled. This can be reinforced by using
appropriate selection rules to suppress the other transitions. The $d-1$
neighboring transitions are driven by near-resonant laser fields that have a
standing-wave configuration along the trap axis,
\begin{eqnarray}
        \lefteqn{{\bf E}(\hat{x},t)}
\nonumber \\*[0.9ex]
        & & =
        \sum_{j=0}^{d-2}\,
        \mbox{\boldmath $\epsilon$}_{j,j+1}\,
        [E_{j,j+1}\,
            e^{-i\alpha_{j,j+1}t} + \mbox{c.c.}]\,
        \cos(k_{j,j+1}\hat{x} + \varphi)\ \ \ \
\nonumber \\*[-1.5ex]
        & &
\label{field} \\*[-0.9ex]
        & & =
        \sum_{j=0}^{d-2}\,
        \mbox{\boldmath $\epsilon$}_{j,j+1}\,
        [E_{j,j+1}\,
            e^{-i\alpha_{j,j+1}t} + \mbox{c.c.}]
\nonumber \\*[-1.2ex]
        & & \ \ \ \ \times\
        \mbox{\large[}\cos(\varphi) -
            \frac{\eta_{j,j+1}}{\sqrt{q}}
                (\hat{a}^{\dagger} \!+ \hat{a}) \sin(\varphi)
            + {\cal O}(\eta_{j,j+1}^2)\mbox{\large]},
\label{field2}
\end{eqnarray}
where $E_{j,j+1}$ and $\alpha_{j,j+1}$ are the (complex) field amplitudes and
field frequencies corresponding to the atomic transitions, and $\mbox{\boldmath
$\epsilon$}_{j,j+1}$ and $k_{j,j+1}$ are the associated polarizations and wave
vector components. The field dependence on $\hat{y}$ and $\hat{z}$ has been
suppressed due to the strong trap confinement along these directions. When
$\varphi = \pi/2$ or $0$, the standing waves make a node or antinode at the
ion's equilibrium position, $\langle\hat{x}\rangle = 0$. We have used
\begin{equation}\label{positionop}
    \hat{x} \,=\, (\hbar/2qm\nu_x)^{1/2} (\hat{a}^{\dagger} +\, \hat{a})
\end{equation}
for the displacement of the ion from equilibrium, where $qm$ is the effective
mass of the CM mode. Each cosine in Eq.~(\ref{field}) has been expanded in
powers of the corresponding Lamb-Dicke parameter, $\eta_{j,j+1} =
k_{j,j+1}(\hbar/2m\nu_x)^{1/2}$, and only terms up to first order in
$\eta_{j,j+1}$ are kept in the limit, $\eta_{j,j+1} \!\ll\! 1$, for each $j$.
For the level scheme under consideration, the internal dipole moment of the ion
is effectively
\begin{equation}
\label{dipmom}
    \hat{{\bf d}} \,=\, \sum_{j=0}^{d-2} \;
    [{\bf d}_{j,j+1} \hat{\sigma}_{j,j+1}^{\dagger} +\,
    {\bf d}_{j,j+1}^{*} \hat{\sigma}_{j,j+1}],
\end{equation}
where $\hat{\sigma}_{j,j+1}$ and ${\bf d}_{j,j+1}^*$ are the transition
operator and matrix element corresponding to the downward transition between
levels \ket{j} and \ket{j+1}. The ion-field interaction is described in the
dipole approximation, using the Hamiltonian, \vspace{-0.3ex}
\begin{equation}\label{hamiltonianfield}
\hat{H}_{\rm{dip}} = -\hat{{\bf d}} \cdot {\bf E}(\hat{x},t),
\end{equation}
where the field depends on the center-of-mass position, $\hat{x}$, of the ion.
This interaction couples the electronic ($\hat{\sigma},\hat{\sigma}^{\dagger}$)
and vibrational ($\hat{a},\hat{a}^{\dagger}$) degrees of freedom of the ion.

We study the time evolution in the interaction picture, where the operators
$\hat{a}(t)$ and $\hat{\sigma}_{j,j+1}(t)$ evolve according to $H_0$, as
$\hat{a}\, e^{-i \nu_x t}$ and $\hat{\sigma}_{j,j+1}\, e^{-i \/\omega_{j,j+1}
t}$. When $\varphi = 0$, the field expansion in Eq.~(\ref{field2}) does not
contain $\hat{a}^\dagger$ and $\hat{a}$ to first approximation, and
$H_{\rm{dip}}$ only affects the internal states of the ion. Tuning each laser
to resonance in this case, $\alpha_{j,j+1} = \omega_{j,j+1}$, we find that
$H_{\rm{dip}}$ becomes time-independent in the interaction picture under the
rotating-wave approximation,
\begin{equation}
\label{hamiltonianV}
    \hat{H}_{\rm{dip}, V} =
    -\hbar \sum_{j=0}^{d-2} \;
    [\Omega_{j,j+1}\,\hat{\sigma}_{j,j+1}^{\dagger} +
    \Omega_{j,j+1}^{*}\,\hat{\sigma}_{j,j+1}],
\end{equation}
where $\Omega_{j,j+1} = ({\bf d}_{j,j+1} \cdot \mbox{\boldmath
$\epsilon$}_{j,j+1} E_{j,j+1})/\hbar$ is the Rabi frequency for the transition
between levels \ket{j} and \ket{j+1}. Alternately, when $\varphi = \pi/2$, the
field is linear in $\hat{a}^\dagger$ and $\hat{a}$ to first approximation, and
$H_{\rm{dip}}$ affects both the internal and external states of the ion. In
this case, detuning each laser above or below resonance by the trap frequency,
$\alpha_{j,j+1} = \omega_{j,j+1} \pm \nu_x$, we find
\begin{equation}
\label{hamiltonianU+}
    \hat{H}_{\rm{dip}, U_+} =
    \hbar \sum_{j=0}^{d-2} \frac{\eta_{j,j+1}}{\sqrt{q}}\;
    [\Omega_{j,j+1} \hat{\sigma}_{j,j+1}^{\dagger} \hat{a}^{\dagger}
    +\, \Omega_{j,j+1}^{*} \hat{\sigma}_{j,j+1} \hat{a}];
\end{equation}
\vspace{-4ex}
\begin{equation}
\label{hamiltonianU-}
    \hat{H}_{\rm{dip}, U_-} =
    \hbar \sum_{j=0}^{d-2} \frac{\eta_{j,j+1}}{\sqrt{q}}\;
    [\Omega_{j,j+1} \hat{\sigma}_{j,j+1}^{\dagger} \hat{a}
    +\, \Omega_{j,j+1}^{*} \hat{\sigma}_{j,j+1}
        \hat{a}^{\dagger}].
\end{equation}
The unitary time evolution operator corresponding to Eq.~(\ref{hamiltonianV}),
\begin{equation}
\label{Vdef}
    \hat{V}=\exp[-i(t/\hbar)\,\hat{H}_{\rm{dip},V}],
\end{equation}
mixes the $d$ internal states of the ion without affecting the trap state, and
turns out to be sufficient for generating the single-qudit gates $Z_d$, up to
an overall phase factor. The time evolution operators corresponding to
Eqs.~(\ref{hamiltonianU+},\ref{hamiltonianU-}),
\begin{equation}
\label{Udef}
    \hat{U}_\pm = \exp[-i(t/\hbar)\,\hat{H}_{\rm{dip},U_\pm}],
\end{equation}
conditionally couple the internal and external co-ordinates of the ion.
Whenever the internal energy of the ion is raised ($\hat{\sigma}^{\dagger}$),
$\hat{U}_+$ raises the trap energy ($\hat{a}^{\dagger}$), while $\hat{U}_-$
lowers the trap energy ($\hat{a}$), in tandem. This conditionality arises from
the rotating-wave approximation, which retains only the energy-conserving terms
in the Hamiltonian. Using $\hat{U}_\pm$ and $\hat{V}$ in stages, we will show
that the two-qudit gates, ${\bf \Gamma_2}[Y_d]$, can be implemented between two
ions in the trap.

First we show how to construct $Z_d$ from $\hat{V}$. In the binary case, $d=2$,
we set all the Rabi frequencies except $\Omega_{0,1}$ equal to zero. The levels
\ket{0} and \ket{1} then undergo two-level Rabi oscillations,
\begin{equation}
\label{binaryV}
    \hat{V} \;\rightarrow\; \Omega^{-1}
    \left[
    \begin{array}{cc}
        \Omega\, C & i\,\Omega_{0,1}^{*} S \\*
        i\,\Omega_{0,1} S & \Omega\, C
    \end{array}
    \right] \hspace{-0.8ex}
    \begin{array}{l}
        \ket{0} \\* \ket{1}
    \end{array},
\end{equation}
where $C = \cos \Omega t, S = \sin \Omega t$, and $\Omega = |\Omega_{0,1}|$.
Given a state $c_0 \ket{0} + c_1 \ket{1}$, we can choose $\Omega_{0,1}$ and $t$
such that
\begin{equation}\label{binaryZ}
    \ \ \
    \frac{\Omega_{0,1}}{\Omega} = \frac{c_0^* c_1}{i |c_0 c_1|}\,;
    \ \ \ \
    \cos\Omega t =  |c_1|\,,
\end{equation}
which makes $\hat{V}$ implement the transform of Eq.~(\ref{Z2}), up to an
overall phase, $i\arg c_1$, simulating the binary gate $Z_2$. In the ternary
case, we set all the Rabi frequencies except $\Omega_{0,1}$ and $\Omega_{1,2}$
equal to zero, leaving a three-level $\Lambda$-system (see Fig.~\ref{levels}).
In this case, the levels \ket{0}, \ket{1} and \ket{2} evolve according to
\begin{equation}
\label{ternaryV}
    \hspace{-37ex}
    \hat{V} \;\rightarrow\; \Omega^{-2} \;\times
\end{equation}
\vspace{-4ex}
\begin{displaymath}
    \left[ \hspace{-0.5ex}
    \begin{array}{c@{\hspace{2ex}}c@{\hspace{2ex}}c}
        |\Omega_{1,2}|^{2} + |\Omega_{0,1}|^2 C &
            i\, \Omega_{0,1}^{*}\Omega S &
            \Omega_{0,1}^{*} \Omega_{1,2} (C-1) \\*
        i\, \Omega_{0,1} \Omega S &
            \Omega^2 C &
            i\, \Omega_{1,2} \Omega S \\*
        \Omega_{0,1} \Omega_{1,2}^{*} (C-1) &
            i\, \Omega_{1,2}^{*} \Omega S &
            |\Omega_{0,1}|^{2} + |\Omega_{1,2}|^2 C
    \end{array}
    \hspace{-0.9ex} \right] \hspace{-0.8ex}
    \begin{array}{l}
        \ket{0} \\* \ket{1}, \\* \ket{2}
    \end{array}
\end{displaymath}
where $C = \cos \Omega t$, $S = \sin \Omega t$, and $\Omega^2 =
|\Omega_{0,1}|^2 + |\Omega_{1,2}|^2$. Given a state $c_0\ket{0} + c_1\ket{1} +
c_2\ket{2}$, we can choose $\Omega_{0,1}$, $\Omega_{1,2}$ and $t$ such that
\begin{equation}\label{ternaryZ}
    \frac{\Omega_{0,1}}{\Omega} =
        \frac{c_0^*c_1}{i |c_1|^2}\
        \frac{S}{1\!-C}\,;
    \ \ \ \
    \frac{\Omega_{1,2}}{\Omega} =
        \frac{i c_1 c_2^*}{S\,|c_2|}\,;
\end{equation}
\vspace{-4ex}
\begin{displaymath}
    \hspace{-2ex}
    \cos\Omega t = \frac{|c_1|^2}{1-|c_2|} - 1,
\end{displaymath}
which makes $\hat{V}$ implement the transform of Eq.~(\ref{Zd}) for $d=3$, up
to an overall phase, $i\arg c_2$, simulating the ternary gate $Z_3$. In the
$d$-valued case, we require $\hat{V}$ to implement $Z_d$ in Eq.~(\ref{Zd}) for
an arbitrary $d$,
\begin{equation}
\label{dvaluedV1}
    \begin{array}{c}
    \hat{V}(\Omega_{0,1}, \Omega_{1,2}, \ldots, \Omega_{d-2,d-1};\, t):
    \\*[1ex]
    c_0\ket{0} + c_1\ket{1} + \cdots + c_{d-1}\ket{d-1}
    \;\mapsto\; e^{i\phi}\ket{d-1},
    \end{array}
\end{equation}
where $\phi$ has been introduced to allow for an overall phase offset. The
controls in $\hat{V}$ are the $d-1$ complex Rabi frequencies, and the
interaction time. The adjoint of the transform in Eq.~(\ref{dvaluedV1}) can be
written as
\begin{equation}
\label{dvaluedV2}
    \hat{V}^{\dagger} \ket{d-1} \,=\, e^{-i\phi}\
    [c_0\ket{0} + c_1\ket{1} + \cdots + c_{d-1}\ket{d-1}].
\end{equation}
Projecting this equation onto $\bra{p}$ and taking the complex conjugate of
both sides, we get
\begin{equation}
\label{dvaluedZ}
    \langle d-1 |\,\hat{V}| p \rangle
    \,=\, c_p^*\ e^{i\phi},\ \
    p = 0, 1, \ldots, d-1,
\end{equation}
which relates $d$ of the matrix elements of $\hat{V}$ to the coefficients,
$c_0, \ldots, c_{d-1}$. These $d$ equations have to be inverted to find the
controls, $\Omega_{0,1}, \ldots, \Omega_{d-2,d-1}$, and $t$. Analytical
solutions are given for the binary and ternary cases in Eqs.~(\ref{binaryZ})
and (\ref{ternaryZ}). This allows $\hat{V}$ to implement $Z_d$ up to a phase
gate, $X_d(\phi)$.

Two-qudit gates of the form \Gam{2}{d} can be implemented using both $U_\pm$
and $V$ interactions, and an auxiliary manifold of $d$ additional levels in
each ion. To see this, write the original state of the two-ion system in the
form,
\begin{equation}
\label{twoion1}
    \ket{\Psi}_{\rm{C}} \, \ket{\Phi}_{\rm{T}} \, \ket{\underline{0}},
\end{equation}
where $\ket{\Psi}_{\rm{C}}$  is the original control ion state,
$\ket{\Phi}_{\rm{T}}$ is the original target ion state, and
$\ket{\underline{0}}$ is the trap ground state. Applying a $\pi$-pulse of the
$U_{\pm}$ interaction to the control ion, we first transform all of the
computational states in this ion except $\ket{d-1}_{\rm{C}}$ to their auxiliary
counterparts, conditional on exciting the trap to $\ket{\underline{1}}$. We
leave $\ket{d-1}_{\rm{C}} \ket{\underline{0}}$ unaffected by turning off the
corresponding laser in $U_{\pm}$. We then restore the internal state of the
control ion to its original configuration by using a $\pi$-pulse of the $V$
interaction, which does not affect the trap. The entangled state of the system
is then given by
\begin{equation}
\label{twoion2}
    \ket{\Psi_{d-1}}_{\rm{C}} \, \ket{\Phi}_{\rm{T}} \, \ket{\underline{0}} \,+\,
    \ket{\Psi_{\rm{other}}}_{\rm{C}} \, \ket{\Phi}_{\rm{T}} \, \ket{\underline{1}},
\end{equation}
where $\ket{\Psi}_{\rm{C}} =\ket{\Psi_{d-1}}_{\rm{C}} +
\ket{\Psi_{\rm{other}}}_{\rm{C}} $. We see that all of the control states
except $\ket{d-1}_{\rm{C}}$ are entangled with the trap state
$\ket{\underline{1}}$. Applying a $\pi$-pulse of the $U_\pm$ interaction to the
target ion now, we transform all of its computational states to their auxiliary
counterparts, conditional on de-exciting the trap,
\begin{equation}
\label{twoion3}
    \ket{\Psi_{d-1}}_{\rm{C}} \, \ket{\Phi}_{\rm{T}} \, \ket{\underline{0}} \,+\,
    \ket{\Psi_{\rm{other}}}_{\rm{C}} \, \ket{\Phi_{\rm{aux}}}_{\rm{T}} \,
        \ket{\underline{0}},
\end{equation}
where $\ket{\Phi_{\rm{aux}}}_{\rm{T}}$ is the original target ion state written
in the auxiliary basis. The first term in expression~(\ref{twoion2}) is not
affected by this operation since the trap ground state cannot be de-excited.
Next, applying $\hat{V}$ in the computational basis of the target ion, we
simulate $Y_d = Z_d$ or $X_d$, transforming $\ket{\Phi}_{\rm{T}}$ but not
affecting $\ket{\Phi_{\rm{aux}}}_{\rm{T}}$,
\begin{equation}
\label{twoion4}
    \ket{\Psi_{d-1}}_{\rm{C}} \,\{\hat{Y}_d \ket{\Phi}_{\rm{T}}\} \;
        \ket{\underline{0}} \,+\,
    \ket{\Psi_{\rm{other}}}_{\rm{C}} \, \ket{\Phi_{\rm{aux}}}_{\rm{T}} \,
        \ket{\underline{0}}.
\end{equation}
The target ion state $\ket{\Phi_{\rm{aux}}}_{\rm{T}}$ is then restored to the
computational basis by reversing the operations that took us from
expression~(\ref{twoion1}) to expression~(\ref{twoion3}), giving
\begin{equation}
\label{twoion5}
    \ket{\Psi_{d-1}}_{\rm{C}} \,\{\hat{Y}_d \ket{\Phi}_{\rm{T}}\} \;
        \ket{\underline{0}} \,+\,
    \ket{\Psi_{\rm{other}}}_{\rm{C}} \, \ket{\Phi}_{\rm{T}} \,
        \ket{\underline{0}}.
\end{equation}
This completes the implementation of \Gam{2}{d} on expression~(\ref{twoion1}),
with the target ion transformed by $Y_d$ conditional on the control ion being
in $\ket{d-1}_{\rm{C}}$. This two-qudit logic is made possible by the
$\hat{U}_\pm$ interaction, which allows the information about whether the
control ion is in $\ket{\Psi_{d-1}}_{\rm{C}}$ or
$\ket{\Psi_{\rm{other}}}_{\rm{C}}$ to be carried to the target ion via
entanglement with the trap states, \ket{\underline{0}} or \ket{\underline{1}}.
This shows that universal multi-valued computing is feasible in the linear ion
trap scheme.

\section{Summary}\label{sec-Conclusion}

We conclude with a comparison of the binary and multi-valued approaches to
quantum computing. The main advantage of the latter is a logarithmic reduction
in the number of separate quantum systems needed to span the quantum memory.
For a Hilbert space of $N$ dimensions, corresponding to $n_2 = \log_2 N$
qubits, the number of qudits needed to store this information is
\begin{equation} \label{spacecomplex}
    n = \frac{\log_2 N}{\log_2 d} = \frac{n_2}{\log_2 d}\,.
\end{equation}
Note that this retains the same scaling in $N$ and $d$ with the inclusion of
the auxiliary qudits used in the gate construction of Fig.~\ref{gates}. Using a
binary equivalent of this construction, we find that the overall
time-complexity of a binary simulation is ${\cal O}[n_2^2 N^2]$. That is, this
many two-qubit gates are required to simulate an $N\/$-dimensional unitary
operator ${\bf U}$. By analogy, the number of two-qudit gates used in the
construction of section~\ref{sec-Gates} scales as ${\cal O}[n^2 N^2]$, or
\begin{equation}\label{timecomplex}
    {\cal O}\!\left[n^2 N^2\right] =\,
    {\cal O}\!\left[\frac{(\log_2 N)^2 N^2}{(\log_2 d)^2}\right] =\,
    {\cal O}\!\left[\frac{n_2^2 N^2}{(\log_2 d)^2}\right]\!,
\end{equation}
where we have used Eq.~(\ref{spacecomplex}). Eq.~(\ref{timecomplex}) represents
an upper bound on the time-complexity of a multi-valued simulation, and shows
that this has a $(\log_2 d)^2$ advantage over the binary case. This comes at
the cost of larger elementary gates, \Gam{2}{d}, which require $d$ states in
each ion to be controlled, not just two. This suggests that we ought to
multiply the multi-valued time-complexity in Eq.~(\ref{timecomplex}) by $d$ for
a physically relevant comparison with the binary case. However, the
frequency-domain approach taken to constructing the $V$ and $U_\pm$
interactions in Eqs.~(\ref{hamiltonianV}-\ref{hamiltonianU-}) assumes that the
$d-1$ lasers operate simultaneously on each ion, which allows a two-qudit gate
to be implemented without slowing down the computation in real time. The cost
of the multi-valued speed-up in this case is the need for multiple lasers to
address the corresponding transitions in the $d\/$-level ion.

Finally, we must make note of the non-logarithmic scaling in $N$ in
Eq.~(\ref{timecomplex}), which shows the inefficiency of this construction for
simulating arbitrary $N$-dimensional unitary transforms. This is analogous to
the binary case, where only certain unitary transforms admit an efficient
simulation in terms of elementary gates, making them useful for efficient
quantum algorithms.

We thank David Aronstein for helpful comments.  This work was supported by the
Army Research Office through the MURI Center for Quantum Information.
\vspace{-1ex}
%
\makeatletter


\begin{references}
\vspace{-12ex}
\bibitem[*]{email}Electronic Address: amuthuk@optics.rochester.edu
\vspace{2ex}
\bibitem{DiVincenzo95}
D.~P. DiVincenzo, Phys. Rev. A {\bf 51}, 1015 (1995).

\bibitem{Sleator95}
T. Sleator and H. Weinfurter, Phys. Rev. Lett. {\bf 74}, 4087 (1995).

\bibitem{Barenco95a}
A. Barenco, Proc. R. Soc. London. A {\bf 449}, 679 (1995).

\bibitem{Barenco95b}
A. Barenco et al., Phys. Rev. A {\bf 52}, 3457 (1995).

\bibitem{Toffoli80}
T. Toffoli, Tech. Memo MIT/LCS/TM-151, MIT Lab. for Comp. Sci. (1980).

\bibitem{Cirac95}
J.~I. Cirac and P. Zoller, Phys. Rev. Lett. {\bf 74}, 4091 (1995).

\bibitem{Monroe95}
C. Monroe et al., Phys. Rev. Lett. {\bf 75}, 4714 (1995).

\bibitem{Rine}
D.~C. Rine, ed., {\em Computer Science and Multiple-valued Logic} (Elsevier
Science Publishing Company, Inc., New York, 1984).

\bibitem{Chau97}
H.~F. Chau, Phys. Rev. A {\bf 55}, R839 (1997).

\bibitem{Rains99}
E.~M. Rains, I.E.E.E. Trans. on Info. Theory {\bf 45}, 1827 (1999).

\bibitem{Ashikhmin00}
A. Ashikhmin and E. Knill, e-print quant-ph/0005008.

\bibitem{Gottesman99}
D. Gottesman, Chaos, Solitons and Fractals {\bf 10}, 1749 (1999).

\bibitem{Aharonov99}
D. Aharonov and M. Ben-Or, e-print quant-ph/9906129.

\bibitem{Burlakov99}
A.~V. Burlakov et al., Phys. Rev. A {\bf 60}, R4209 (1999).

\bibitem{Grover97}
L.~K. Grover, Phys. Rev. Lett. {\bf 79}, 325 (1997).

\bibitem{Deutsch89}
D. Deutsch, Proc. R. Soc. London A {\bf 425}, 73 (1989).

\bibitem{Shi90}
S. Shi and H. Rabitz, J. Chem. Phys. {\bf 92}, 364 (1990).

\bibitem{Noel97}
M.~W. Noel and C.~R. Stroud, Jr., Optics Express {\bf 1}, 176 (1997).

\bibitem{Araujo98}
L.~E.~E. de Araujo, I.~A. Walmsley, and C.~R. Stroud, Jr., Phys. Rev. Lett.
{\bf 81}, 955 (1998).

\end{references}
\end{document}